\documentclass[aps,groupedaddress,showpacs,twocolumn]{revtex4}
\usepackage{color,graphicx,amsmath,amssymb,bm,epsfig,citesort}

\begin{document}

\def\bra#1{\left\langle{#1}\,\right|\,}    
\def\ket#1{\,\left|\,{#1}\right\rangle}    

\title{Clauser-Horne inequality and decoherence in mesoscopic conductors}

\author{Elsa Prada}
\affiliation {Departamento de F\'{\i}sica Te\'orica de la Materia
Condensada, C-V, Universidad Aut\'onoma de Madrid, E-28049 Madrid,
Spain\\ Institut f\"{u}r Theoretische Festk\"{o}rperphysik,
Universit\"{a}t Karlsruhe, D-76128 Karlsruhe, Germany}

\author{Fabio Taddei}
\affiliation{NEST-INFM \& Scuola Normale Superiore, I-56126 Pisa, Italy}

\author{Rosario Fazio}
\affiliation{NEST-INFM \& Scuola Normale Superiore, I-56126 Pisa, Italy}

\date{\today}

\begin{abstract}
We analyze the effect of decoherence on the violation of the
Clauser-Horne (CH) inequality for the full electron counting
statistics in a mesoscopic multiterminal conductor. Our setup
consists of an \emph{entangler} that emits a flux of entangled
electrons into two conductors characterized by a scattering matrix
and subject to decoherence.
Loss of phase memory is modeled phenomenologically by introducing fictitious extra leads.
The outgoing electrons are detected
using spin-sensitive electron counters. Given a certain average
number of incoming entangled electrons, the CH inequality is
evaluated as a function of the numbers of detected particles and on
the various quantities characterizing the scattering matrix. When
decoherence is turned on, we show that the amount of violation of
the CH inequality is effectively reduced. Interestingly we find
that, by adjusting the parameters of the system, there exists a
protected region of $Q$ values for which violation holds for
arbitrary strong decoherence.
\end{abstract}


\maketitle

\section{\label{Sec:introduction}Introduction}

Entanglement \cite{bell87} is probably the most important resource
for the implementation of quantum computation and quantum
communication protocols \cite{nielsen00}. Since recently, most of
the work on entanglement has been carried out in optical systems
using photons \cite{zeilinger99}, cavity QED systems
\cite{rauschenbeutel00}, and ion traps \cite{sackett00}. Solid state
systems, however, are a very attracting arena of research in quantum
information \cite{loss98,makhlin01,awschalom02} because, in
perspective of future applications, they should allow for
scalability and integration. In this light, a number of different
realizations of entangled electrons have since been proposed: hybrid
normal-superconducting structures
\cite{loss01,lesovik01,chtchelkatchev02,samuelsson03,recher03,prada04,samuelsson04-2,sauret04},
superconductor-carbon nanotubes systems
\cite{recher02,bena01,bouchiat03}, quantum dots in the Coulomb
blockade regime \cite{loss98,oliver02,saraga03}, chaotic quantum
dots \cite{beenakker03-2}, Kondo-like impurities \cite{costa01},
quantum Hall bar systems
\cite{beenakker03,samuelsson04,beenakker04,beenakker04-2,samuelsson04-3},
Coulomb scattering in 2D electron gas \cite{saraga04}.

Besides its generation, a crucial issue is that of the detection of
entanglement. By means of a beam-splitter, entanglement can be
detected in transport through an analysis of current noise
\cite{burkard00} or higher cumulants \cite{taddei02}. Furthermore,
the presence of entanglement can be revealed by analyzing the Bell
inequality and quantities like concurrence \cite{wootters98}, which
have been expressed in terms of zero-frequency charge and
spin-current noise
\cite{kawabata01,lesovik01,chtchelkatchev02,samuelsson03,beenakker03,beenakker04-2,sauret04B}.
Violation of a Bell inequality implies that there exist quantum
correlations between the detected particles that cannot be described
by any local hidden variables theory. In the same spirit as it was
done for the noise, in Ref.~\onlinecite{faoro04} a Clauser-Horne(CH)
inequality \cite{clauser74,mandel95} was derived for the Full
Counting Statistics (FCS) of electrons and its properties were
discussed \cite{newnota2}. In particular, it was found that the
maximum violation of the CH inequality for electrons in the Bell
state simply scales as the inverse of the number of injected
particles. It was also found that the CH inequality is violated for
a superconducting hybrid structure and, more interestingly, for a
three terminal fully normal device.

In real systems electrons are unavoidably coupled to the
electromagnetic environment. As a result dephasing takes place,
thereby reducing and eventually destroying entanglement.
Understanding the consequences of dephasing is an important issue.
In Refs.
\onlinecite{samuelsson03,samuelsson03-2,samuelsson04,samuelsson04-2}
the effect of dephasing was mimicked by introducing in the density
matrix of the electronic entangled states a phenomenological
parameter which suppresses its off-diagonal elements. By properly
choosing the transmission probability of beam-splitters or tunnel
barriers, violation of Bell inequality was found even for ``strong''
dephasing. In Refs. \onlinecite{beenakker03,beenakker03-3} dephasing
was introduced averaging over an uniform distribution of random
phase factors accumulated in each edge channel of the quantum Hall
bar. If the two edge channels are mixed by the tunnel barrier, no
violation was reported for ``strong'' dephasing. The effect of
decoherence and relaxation has also been analyzed using a Bloch
equation formalism in Ref. \onlinecite{burkard03}.

In the present work we analyze the CH inequality for the FCS
\cite{faoro04} in the presence of dephasing. We consider the
prototype setup depicted in Fig.~\ref{2VP}, consisting of a generic
entangler connected to two conducting wires. Entangled electrons
injected in the two leads are detected by performing spin-selective
counting along a given local quantization axis. The entangled
electrons are subject to decoherence while transversing the
conductors (thus before reaching the detectors)\cite{newnota1}.
Various phenomenological methods have been developed to treat
dephasing in transport through mesoscopic conductors. In Refs.
\onlinecite{seelig01,marquardt04}, which actually describes exactly
nonequilibrium radiation acting on the system, dephasing is induced
by a classical fluctuating potential. In Ref. \onlinecite{pala04},
dephasing is treated as random fluctuations of the phase of
propagating modes through the conductor. Both methods have been
recently applied to FCS in Refs. \onlinecite{pala04-2,forster05}. In
this paper decoherence is introduced as due to the presence of
additional fictitious reservoirs along both wires. This method,
which mimics the effect of inelastic processes, was introduced by
B\"uttiker \cite{buttiker86,buttiker88} in terms of fictitious extra
leads \cite{nota1}. The advantage of this model resides in the fact
that inelastic, phase randomization processes are implemented within
an elastic, time-independent scattering problem. In the rest of the
paper we shall refer to decoherence as to the effect produced by
such fictitious additional leads.

As expected, we find that decoherence suppresses the violation of
the CH inequality, though leaving unchanged the range of angles for
which violation occurs. In particular, the value of the maximum
violation is suppressed more rapidly as compared with the absence of
decoherence (exponentially with the square root of the number of
injected electrons instead of algebraically). Importantly, by
studying the CH inequality as a function of the number of
transmitted electrons, there exist values of such quantity that are
more protected against decoherence.

The paper is organized as follows: In Section \ref{Sec:system} we
described in detail the mesoscopic system we are considering to test
the violation of the CH inequality together with the
phenomenological model of decoherence. Section \ref{Sec:CHI} is
devoted to the formulation of the CH inequality for the FCS within
the scattering approach and to the analysis of the no-enhancement
assumption (Section \ref{Subsec:noen}). The results are presented in
Section \ref{sec:results}, where a systematic analysis of the
violation of the CH inequality against all the parameters of the
device is addressed. A concluding summary is provided in Section
\ref{Sec:conclusions}.

For completeness, we include in Appendix \ref{App:asym} the results
relative to an asymmetric setup, whereby decoherence occurs only in
one of the two wires. In Appendix \ref{Ap:expecvalues} and
\ref{Ap:probdistributions} we collect, respectively, the expressions
of the expectation values and the different probability
distributions.

\section{\label{Sec:system}Description of the system}

\begin{figure}
\includegraphics[width=7cm]{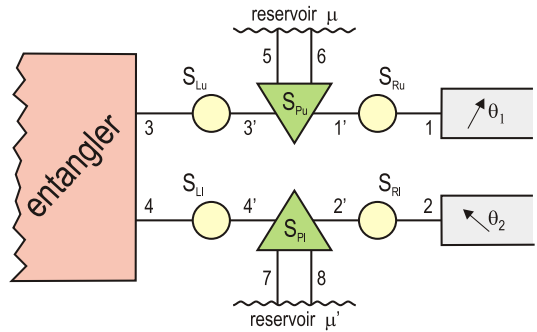}
\caption{(Color online) Idealized setup for testing the CH
inequality for electrons in a solid-state environment in the
presence of decoherence. It consists of three parts: An entangler
that produces pairs of spin-entangled electrons exiting from
terminals 3 and 4. Two conductors that connect these terminals with
exiting leads 1 and 2, and two analyzers. The conductors are
described by the elastic scattering matrices $S_{\text{Lu}},
S_{\text{Ru}}, S_{\text{Ll}}$ and $S_{\text{Rl}}$, and the inelastic
ones $S_{\text{Pu}}$ and $S_{\text{Pl}}$. These last ones can
simulate phenomenologically the presence of decoherence through the
coupling via two leads ($5, 6$ and $7, 8$) to two additional
reservoirs of chemical potentials $\mu$ and $\mu'$. Electron
counting is performed in leads 1 and 2. Finally, $\theta_1$ and
$\theta_2$ are the angles at which the spin-quantization axis are
oriented.}\label{2VP}
\end{figure}

We consider the setup illustrated in Fig.~\ref{2VP}. It consists of
an \emph{entangler}, two conducting wires and two spin-selective
counters. The entangler, on the left-hand-side, is a device that
produces pairs of electrons, with energy $E<\mu_{\textrm{L}}$, in a
maximally spin entangled state (Bell state). On the right-hand-side
of Fig.~\ref{2VP} the electron counting is performed in leads 1 and
2 (at equal electrochemical potential $\mu_{\textrm{R}}$) for
electrons with spin aligned along the local spin-quantization axis
at angles $\theta_1$ and $\theta_2$ (spin-selective counters). As a
convention we say that the analyzer is not present when the electron
counting is spin-insensitive (electrons are counted irrespective of
their spin direction). Since we assume no back-scattering from
counters to the entangler, the particles which are not counted are
lost and hence there is no communication between the two detectors.
Leads 3 and 4 of the entangler are connected to exit leads 1 and 2
through two conductors, where inelastic processes are introduced
through the fictitious lead model of B\"uttiker \cite{buttiker84}. Let
us analyze in detail the upper wire (see Fig.~\ref{2VP}) which
connects the emitting lead $3$ with the exiting lead $1$. The
conductor consists of three scattering regions. The elastic
scatterer connecting lead 3 to 3' is described by the matrix
\begin{equation}
\hat{S}_{\text{Lu}}=\left(\begin{array}{cc}
\check{r} & \check{t'} \\
\check{t} & \check{r'}\end{array} \right )= \left
(\begin{array}{cccc}
r_{\uparrow} & 0 & t'_{\uparrow} & 0 \\
0 &  r_{\downarrow} & 0 &  t'_{\downarrow} \\
t_{\uparrow} & 0 & r'_{\uparrow} & 0 \\
0 &  t_{\downarrow} & 0 &  r'_{\downarrow}
\end{array}\right) .
\end{equation}
[The index L (R) stands for Left (Right) elastic scatterer, while P
stands for Probe scatterer; u (l) refers to the upper (lower) wire.]
Here $r_{\sigma}$ ($t_{\sigma}$) is the probability amplitude for an
incoming particle with spin $\sigma$ from lead 3 to be reflected
(transmitted into lead 3'). For a normal single-channel wire we set
$t_{\sigma}=t'_{\sigma}=\sqrt{T_0}$, and
$r_{\sigma}=r'_{\sigma}=i\sqrt{1-T_0}$, where $T_0$ is the
transmission probability. Inelastic scattering is introduced by
plugging in an additional reservoir of chemical potential $\mu$ with
an energy- and spin-independent scattering matrix
\begin{equation}\label{Salpha}
\bar{S}_{\text{Pu}}=\left(\begin{array}{cccc}
\check{0} & \sqrt{1-\alpha}~\check{1} & \sqrt{\alpha}~\check{1} &  \check{0} \\
\sqrt{1-\alpha}~\check{1} & \check{0} & \check{0} & \sqrt{\alpha}~\check{1} \\
\sqrt{\alpha}~\check{1} & \check{0} & \check{0} & -\sqrt{1-\alpha}~\check{1} \\
\check{0} & \sqrt{\alpha}~\check{1} & -\sqrt{1-\alpha}~\check{1} & \check{0}
\end{array} \right),
\end{equation}
represented by a triangle in Fig.~\ref{2VP}. For the sake of
clarity, we have denoted by a check ($\check{~}$), a caret
($\hat{~}$) and an overbar ($\bar{~}$), respectively, (2$\times$2),
(4$\times$4) and (8$\times$8) matrices. In Eq.~(\ref{Salpha})
$\check{1}$ and $\check{0}$ are, respectively, unit and zero
(2$\times$2)-matrices in the spin space, and $(1-\alpha)$ is the
probability for transmitting a particle between leads 3' and 1'. The
coupling parameter ranges from $\alpha=0$, when no particles are
transmitted into leads 5 and 6 from leads 3' and 1', to $\alpha=1$,
when no particles are transferred between leads 3' and 1'. A third
elastic scatterer, described by a matrix $\hat{S}_{\text{Ru}}$,
connects lead 1' to lead 1. The conductor is therefore described by
the matrix $\hat{S}_{13}$ defined as
$\hat{S}_{13}=\hat{S}_{\text{Ru}}\otimes \bar{S}_{\text{Pu}}\otimes
\hat{S}_{\text{Lu}}$, where the notation $\otimes$ stands for the
scattering matrix composition (elimination of internal current
amplitudes) \cite{datta95}. For simplicity, we shall assume that
$\hat{S}_{\text{Ru}}=\hat{S}_{\text{Lu}}$.

Due to the presence of the additional reservoir,
particles propagating through lead 3 are transmitted partially to
lead 1 and partially to leads 5 and 6 (see Fig.~\ref{2VP}). The
additional reservoir, however, can transfer itself particles to lead
1 and 3. As a result, only a fraction of the particles arriving in
lead 1 comes from {\em coherently} transmitted ones sent in from
lead 3, with probability
\begin{equation}
T_{13}=\frac{T_0^2(1-\alpha)}{[1+(1-T_0)(1-\alpha)]^2}.
\end{equation}
Another fraction, the {\em incoherent} contribution, comes from the
additional reservoir through leads 5 and 6, with probability
\begin{equation}
T_{15}+T_{16}=\frac{T_0\alpha}{1+(1-T_0)(1-\alpha)}.
\end{equation}
The presence of the extra reservoir mimics the fact that the current
flowing through the conductor is partially composed of particles
(the {\em incoherent}  fraction) which have lost phase memory while
traversing it. For $\alpha=0$ all particles are coherently
transmitted and $T_{15}+T_{16}=0$, while for $\alpha=1$ all
particles are transferred incoherently and $T_{13}=0$. For
$\alpha=0$, the overall transmission probability through the
conductors is given by $T=T_0^2/(2-T_0)^2$. In the rest of the paper
we will refer to $\alpha$ as to the {\em decoherence rate}.

The chemical potential $\mu$ of the additional reservoir is set in
such a way that no net current flows in or out of the reservoir
($I_5+I_6=0$). This constraint is enforced only on average. An
instantaneous current in or out the additional reservoir is then
allowed \cite{buttiker86,buttiker88,beenakker92,blanter00}, and a
non-fluctuating chemical potential $\mu$ is assumed (for this reason
the additional terminal does not behave as a voltage probe).

A similar description applies to the lower wire connecting lead 4
with lead 2, so that the scattering matrix of the conductor is
defined as $\hat{S}_{24}=\hat{S}_{\text{Rl}}\otimes
\bar{S}_{\text{Pl}}\otimes \hat{S}_{\text{Ll}}$, where, for
simplicity, we set $\hat{S}_{\text{Rl}}=\hat{S}_{\text{Ll}}$. If the
angles $\theta_1$ and $\theta_2$ of the analyzers are parallel to
each other and in the absence of spin mixing processes, the total
matrix of the system can be written as
\begin{equation}\label{S-ent}
\bar{S}=\left(\begin{array}{cc}
\hat{S}_{13} & \hat{0}\\
\hat{0} & \hat{S}_{24}
\end{array}\right)\,.
\end{equation}
The general scattering matrix relative to non-collinear angles
$\bar{S}_{\theta_1,\theta_2}$ is obtained from $\bar{S}$ by rotating
the spin quantization axis independently in the two conductors (note
that this is possible because the two wires are decoupled) \cite{brataas01}:
$\bar{S}_{\theta_1,\theta_2}=\bar{{\cal U}} \bar{S}
\bar{{\cal U}}^{\dagger}$, where $\bar{{\cal U}}$ is the rotation
matrix given by
\begin{equation}\label{tra}
\bar{{\cal U}}=\left( \begin{array}{cccc}
\check{1} & \check{0} & \check{0} & \check{0}\\
\check{0} & \check{U}_{\theta_1} & \check{0} & \check{0}\\
\check{0} & \check{0} & \check{1} & \check{0}\\
\check{0} & \check{0} & \check{0} &
\check{U}_{\theta_2}\end{array}\right),
\end{equation}
and
\begin{equation}\label{uteta}
\check{U}_{\theta}=\left( \begin{array}{cc} \cos{\frac{\theta}{2}}&\sin{\frac{\theta}{2}}\\
-\sin{\frac{\theta}{2}}&\cos{\frac{\theta}{2}}
\end{array} \right).
\end{equation}

For simplicity, we further assume that the two conductors are equal
and that they are subjected to the same degree of decoherence, so
that $\hat{S}_{13}=\hat{S}_{24}$. For this reason the chemical
potentials of the additional reservoirs are identical. It is
interesting to notice that decoherence processes in the two wires
are, in some sense, "uncorrelated", meaning that we have imposed
that the currents flowing through the fictitious leads vanish
separately in the two reservoirs.
(Correlations can be introduced, for example, by imposing $I_5+I_6+I_7+I_8=0$.) In the symmetrical
setup we are considering here,
$\mu=\mu_{\textrm{R}}+(\mu_{\textrm{L}}-\mu_{\textrm{R}})/2$. In the
rest of the paper we consider the case in which all the reservoirs
are at zero temperature.

The incoming state of the system $\ket{\psi}$ depends on whether the
energy of electrons falls within the range $\mu_{\textrm{R}}<E<\mu$
or $\mu<E<\mu_{\textrm{L}}$:
\begin{equation}\label{st}
\ket{\psi}=\left\{
\begin{array}{lc}
\ket{\psi_{\textrm{B}}} & ~~~~~\mu<E<\mu_{\textrm{L}} \\
\ket{\psi_{\textrm{S}}} & ~~~~~\mu_{\textrm{R}}<E<\mu
\end{array}\right. ,
\end{equation}
where
\begin{equation}\label{stB}
\ket{\psi_{\textrm{B}}}=\frac{1}{\sqrt{2}}
\left[a^{\dagger}_{3\uparrow}(E) a^{\dagger}_{4\downarrow}(E)\pm
a^{\dagger}_{3\downarrow}(E) a^{\dagger}_{4\uparrow}
(E)\right]\ket{0},
\end{equation}
and
\begin{eqnarray}\label{stS}
\ket{\psi_{\textrm{S}}}&=&\frac{1}{\sqrt{2}}\left[a^{\dagger}_{3\uparrow}(E)
a^{\dagger}_{4\downarrow}(E)\pm  a^{\dagger}_{3\downarrow}(E)
a^{\dagger}_{4\uparrow}(E)\right]\nonumber\\
&\times&\prod_{n=5,6,7,8}
a^{\dagger}_{n\uparrow}(E)a^{\dagger}_{n\downarrow}(E)\ket{0} .
\end{eqnarray}
In  Eqs.~(\ref{stB}) and (\ref{stS}) $a^{\dagger}_{i\sigma}(E)$ is
the creation operator for a propagating electron in lead $i$ with
spin $\sigma$ at energy $E$. The upper sign refers to the case in
which the incoming state is a spin triplet and the lower sign to the
spin singlet. Electrons with energy between $\mu$ and
$\mu_{\textrm{L}}$ are exiting leads $3$ and $4$ of the entangler in
a superposition of spin $\uparrow$ and $\downarrow$ states. For
energies between $\mu_{\textrm{R}}$ and $\mu$ electrons are also
injected from the additional leads (with indexes 5, 6, 7 and 8) in a
factorized state. Note that this occurs only in the presence of
decoherence, i.e. for $\alpha\neq 0$.

By setting $\mu_{\textrm{R}}=0$ and $\mu_{\textrm{L}}=eV$, the total
current flowing in lead 1, calculated using the Landauer-B\"uttiker
formalism \cite{landauer57} in the linear response regime, is given
by
\begin{equation}
I_1=e^2 V/h(T_{13}+T_{15}+T_{16}). \label{i1}
\end{equation}
Although the coherent part of the current decreases with $\alpha$,
the total current increases with it (except for $T=1$, where it
remains constant and equal to $e^2 V/h$). We would like to mention
that this is a special feature of the model of decoherence we are
using, not to be expected in general.

\section{\label{Sec:CHI}CH Inequality for the Full Counting Statistics}

The quantity employed in the formulation of the CH inequality, as
derived in Ref. \onlinecite{faoro04}, is the joint probability
$P(Q_1,Q_2)$ for transferring a number of $Q_1$ and $Q_2$ electronic
charges into leads $1$ and $2$ over an observation time $t$. The CH
inequality  is based on the hypothesis that the outcome of a
measurement could be accounted for by a local hidden variable
theory. The test of the CH inequality proceeds as follows. The
entangler is switched on during an observation time $t$ (where the
minimum $t$ is the inverse of the measuring device bandwidth) in
which it emits an average number $M$ of pairs of entangled
electrons. After traversing the conductors (and being affected by
inelastic scattering) the electrons are counted in both terminals 1
and 2. The experiment is then repeated to get single terminal and
joint terminal probability distributions that $Q_1$ particles arrive
into analyzer 1 and $Q_2$ particles arrive into analyzer 2 (along a
local spin-quantization axis or independently of it) with
$Q_1+Q_2\leq 2M$.

The CH inequality for the FCS reads \cite{faoro04}
\begin{eqnarray}\label{CHI}
{\cal
S}_{\textrm{CH}}&=&P^{\theta_1,\theta_2}(Q_1,Q_2)-P^{\theta_1,\theta'_2}(Q_1,Q_2)+
P^{\theta'_1,\theta_2}(Q_1,Q_2)\nonumber\\&&+
P^{\theta'_1,\theta'_2}(Q_1,Q_2)-P^{\theta'_1,-}(Q_1,Q_2)\nonumber\\&&-P^{-,\theta_2}(Q_1,Q_2)\le
0 .
\end{eqnarray}
The possible violation, or
the extent of it, also depends on $Q_1$ and $Q_2$.
$P^{\theta_1,\theta_2}(Q_{1},Q_{2})$ is the joint probability in the
presence of two analyzers, where $Q_{1}$ electrons are counted in
lead 1 along $\theta_1$ direction and $Q_{2}$ are counted in lead 2
along $\theta_2$. $P^{\theta_1,-}(Q_{1},Q_2)$ is the corresponding
joint probability when one of the two analyzers has been removed.
The same notation will be used for single terminal probability
distributions: $P^{\theta_i}(Q_{i})$ in the presence of an analyzer
and $P(Q_{i})$ if no analyzer is present. Eq.~(\ref{CHI})
holds for all values of $Q_1$ and $Q_2$ which satisfy the {\em
no-enhancement assumption}:
\begin{equation}
P^{\theta_i}(Q_{i})\le P(Q_{i}) . \label{noenhancement}
\end{equation}

The joint probability distribution for transferring $Q_{1\sigma}$
electrons with spin $\sigma$ in lead 1, $Q_{2\sigma}$ electrons with spin $\sigma$ in lead 2 and so on is given by
\begin{eqnarray}\label{counting}
&&P(Q_{1\uparrow},Q_{1\downarrow},Q_{2\uparrow},\ldots)\nonumber\\&&=\frac{1}{(2\pi)^{2n}}
\int_{-\pi}^{+\pi} d\lambda_{1\uparrow} d\lambda_{1\downarrow}
d\lambda_{2\uparrow} \ldots ~\chi(
\vec{\lambda_{\uparrow}},\vec{\lambda_{\downarrow}})\nonumber\\
&&\times~e^{i\vec{\lambda_{\uparrow}}\cdot\vec{Q_{\uparrow}}}
~e^{i\vec{\lambda_{\downarrow}}\cdot\vec{Q_{\downarrow}}} ,
\end{eqnarray}
where $\chi(\vec{\lambda_{\uparrow}},\vec{\lambda_{\downarrow}} )$
is its characteristic function that can be expressed within the
scattering approach.

For long measurement times $t$, the total characteristic function
$\chi$ is the product of contributions from different energies, so
that
\begin{equation}\label{chidef}
\chi(\vec{\lambda_{\uparrow}},\vec{\lambda_{\downarrow}})=e^{\frac{t}{h}\int
dE~\log{\chi_E(\vec{\lambda_{\uparrow}},\vec{\lambda_{\downarrow}})}} .
\end{equation}
The energy-resolved characteristic function for the transfer of
particles at a given energy $E$ in a structure attached to $n$ leads
can be written as \cite{muzykanskii94,levitov93,levitov96}
\begin{eqnarray}\label{chiEdef}
\chi_E(\vec{\lambda_{\uparrow}},\vec{\lambda_{\downarrow}})=
\left\langle\prod_{j=1,n} e^{i\lambda_{j\uparrow}
\hat{N}_I^{j\uparrow}}e^{\lambda_{j\downarrow}
\hat{N}_I^{j\downarrow}}\right.\nonumber\\\left.\times\prod_{j=1,n}
e^{-i\lambda_{j\uparrow}
\hat{N}_O^{j\uparrow}}e^{{-i\lambda_{j\downarrow}
\hat{N}_O^{j\downarrow}}}\right\rangle\,,&
\end{eqnarray}
where the brackets $\langle ... \rangle$ stand for the quantum
statistical average over the thermal distributions in the leads.
Assuming a single channel per lead, $\hat{N}_{O(I)}^{j\sigma}$ is
the number operator for outgoing (incoming) particles with spin
$\sigma$ in lead $j$ and $\vec{\lambda_{\uparrow}}$,
$\vec{\lambda_{\downarrow}}$ are vectors of $n$ real numbers, one
for each open channel. In terms of outgoing (incoming) creation
operator $\hat{\phi}_{j\sigma}^{\dagger}$
($\hat{a}_{j\sigma}^{\dagger}$), which are linked by the total
scattering matrix of the system $S$, the number operators can be
expressed as
\begin{equation}
\hat{N}_I^{j\sigma}= \hat{a}_{j\sigma}^{\dagger}\hat{a}_{j\sigma};
\qquad\hat{N}_O^{j\sigma}=\hat{\phi}_{j\sigma}^{\dagger}\hat{\phi}_{j\sigma}.
\end{equation}

At zero temperature, the statistical average over the Fermi
distribution function in Eq.~(\ref{chiEdef}) simplifies to the
expectation value calculated over the state $\ket{\psi}$ defined in
Eq.~(\ref{st}). The interval of integration in Eq.~(\ref{chidef})
can be separated in two energy ranges, namely $E<\mu$ and
$\mu<E<eV$. Since, in the limit of a small voltage bias $V$,
$\chi_E$ is energy-independent, Eq.~(\ref{chidef}) can be
approximated to
\begin{equation}\label{chiapproxSB}
\chi(\vec{\lambda_{\uparrow}},\vec{\lambda_{\downarrow}})\simeq
\left[\chi_0^{\textrm{S}}(\vec{\lambda_{\uparrow}},\vec{\lambda_{\downarrow}})\right]^
{M_{\mu}}\left[
\chi_0^{\textrm{B}}(\vec{\lambda_{\uparrow}},\vec{\lambda_{\downarrow}})\right]^
{M-M_{\mu}} ,
\end{equation}
where $M_{\mu}=\mu t/h$ and $M=eVt/h$.

According to Eq.~(\ref{counting}), both single terminal and joint
probability distributions require the computation of
multidimensional integrals, which can only be performed numerically.
In Appendix \ref{Ap:probdistributions} it is shown that the various
probability distributions needed to evaluate the CH inequality can
be expressed in a differential form, more suitable for numerical
evaluation. All the expectation values needed for the calculations
are collected in Appendix \ref{Ap:expecvalues}. Since the two wires
are decoupled and there are no spin-flip processes, the joint
probabilities with a single analyzer are factorized:
\begin{eqnarray}\label{factorized}
P^{\theta_1,-}(Q_1,Q_2)=P^{\theta_1}(Q_1) P(Q_2),\nonumber\\
P^{-,\theta_2}(Q_1,Q_2)=P(Q_1) P^{\theta_2}(Q_2) .
\end{eqnarray}

Rotational invariance makes $P^{\theta_1,-}(Q_1,Q_2)$ and
$P^{-,\theta_2}(Q_1,Q_2)$ independent of the angle of the analyzers,
while $P^{\theta_1,\theta_2}(Q_1,Q_2)$ depends on the angles only
through the combination $\frac{\theta_1\pm \theta_2}{2}$ (upper sign
for triplet and lower sign for singlet), so that we can define
$P_{1,2}^{\frac{\theta_1\pm\theta_2}{2}}(Q_1,Q_2)\equiv
P^{\theta_1,\theta_2}(Q_1,Q_2)$ and $P_{1,-}(Q_1,Q_2)\equiv
P^{\theta_1,-}(Q_1,Q_2)$. As a result, the CH inequality depends
only on three angles $\theta_a\equiv\theta_1\pm\theta_2$,
$\theta_b\equiv\theta_2\pm\theta'_1$ and
$\theta_c\equiv\theta'_1\pm\theta'_2$
($\theta_d=\theta_1\pm\theta'_2$ is a linear combination of the
other three: $\theta_d=\theta_a+\theta_b+\theta_c$). Since
$P^{\theta_1,\theta_2}(Q_1,Q_2)$ is an even function of
$\frac{\theta_1\pm \theta_2}{2}$, in order to find maximal
violations we can restrict the evaluation of the CH inequality to
the following set of angles:
$\theta_a=\theta_b=\theta_c=\theta_d/3\equiv2\Theta$. (This is found
by imposing that positive contributions to ${\cal S}_{\textrm{CH}}$
are maximum while negative contributions are minimum.) The quantity
${\cal S}_{\textrm{CH}}$, characterizing the CH inequality, will
therefore depend on a single angle $\Theta$, on the decoherence
strength $\alpha$ and on the value of the transmitted charge $Q_1$
and $Q_2$. As a result, the CH inequality takes the simplified form
\begin{eqnarray}\label{CHIreduced}
{\cal
S}_{\textrm{CH}}&=&3P_{1,2}^{\Theta}(Q_1,Q_2)-P_{1,2}^{3\Theta}(Q_1,Q_2)-P_{1,-}(Q_1,Q_2)\nonumber\\&&-P_{-,2}(Q_1,Q_2)\le
0.
\end{eqnarray}
Without loss of generality we can choose $\sigma=\sigma'=\uparrow$.
The other cases can be recovered by rotating the polarizers an angle
$\pi$.

\subsection{\label{Subsec:noen}No-enhancement assumption}

\begin{figure}
\includegraphics[width=8.5cm]{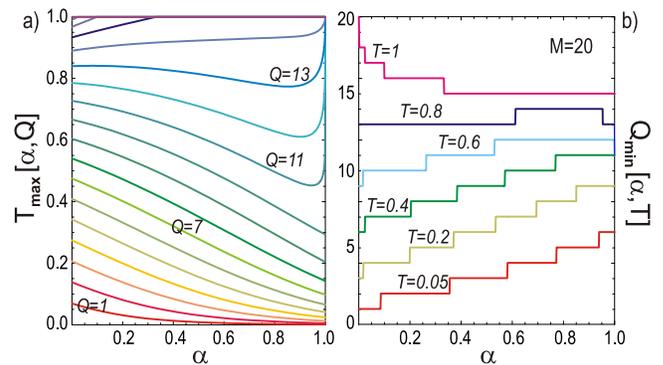}
\caption{(Color online) a) Maximum value of the transmission,
$T_{\text{max}}$, allowed by the no-enhancement assumption as a
function of decoherence rate $\alpha$ for $M=20$ emitted pairs and
for the different values of $Q$. b) Minimum allowed number of
transmitted particles $Q_{\text{min}}$ for a fixed wire transmission
as a function of the decoherence rate. } \label{TmaxQmin20}
\end{figure}

As mentioned above, the CH inequality can be derived under the
no-enhancement assumption,  Eq.~(\ref{noenhancement}). Such a
condition is trivially true when a single particle is transmitted,
$Q=1$: The presence of an analyzer can only decrease the counting
probability \cite{clauser74}. However, when many particles are
transmitted, $Q>1$, the no-enhancement assumption is a relationship
between distribution probabilities that is not, in general,
satisfied for all values of $Q$.

We remind that in the absence of decoherence \cite{faoro04}, for
given $M$ and $Q$, the no-enhancement assumption in one of the two
leads is satisfied only within a range of values of $T$ below
certain threshold $T_{\text{max}}(M,Q)$. In the case of different
numbers of transmitted particles in lead 1 and 2 ($Q_1\neq Q_2$),
the maximum allowed transmission probability must be taken to be the
minimum between $T_{\text{max}}(M,Q_1)$ and $T_{\text{max}}(M,Q_2)$,
according to our assumption of identical wires.

The no-enhancement assumption is affected by decoherence as a
consequence of the fact that single terminal probabilities, with or
without analyzer, depend on $\alpha$. More precisely, the
no-enhancement assumption in one of the two leads is satisfied for
transmissions up to a threshold value which is now a function of
$\alpha$: $T_{\text{max}}(\alpha,M,Q)$. Unlike the ideal case, for
$\alpha\neq 0$ it is not possible to find an analytical expression
for $T_{\text{max}}$. In Fig.~\ref{TmaxQmin20}a $T_{\text{max}}$ is
plotted as a function of $\alpha$ for $M=20$ and all values of $Q$
from 1 to 20. One can see that $T_{\text{max}}$ monotonically
decreases with $\alpha$ for values of $Q\lesssim M/2$ and
monotonically increases for large values of $Q$. For intermediate
values of $Q$, $T_{\text{max}}$ decreases up to values of $\alpha$
close to one and then rapidly increases reaching one when
$\alpha=1$. This behavior is specific of the fictitious lead model
and reflects the fact that both the average total current
[Eq.~(\ref{i1})], related to $P(Q)$, and the average spin-polarized
current, related to $P^{\theta}(Q)$, are increasing functions of
$\alpha$. Indeed, as a consequence of a finite $\alpha$, the two
distributions shift to larger values of $Q$, as it would happen for
an enhanced effective transmission probability. Its maximum allowed
value by the no enhancement assumption is therefore reached for a
smaller $T$. As a consequence $T_{\text{max}}$ must decrease with
$\alpha$. This argument is not valid when $T_{\text{max}}\simeq 1$
at $\alpha=0$, since the average currents do not change appreciably
with $\alpha$ and only the peculiar shape of the distributions
matters. We define
\begin{eqnarray}
&&T_{\text{max}}(\alpha,M,Q_1,Q_2)\nonumber\\&&=\textrm{Min}
[T_{\text{max}}(\alpha,M,Q_1),T_{\text{max}}(\alpha,M,Q_2)] .
\end{eqnarray}

Alternatively, given a wire with a fixed transmission $T$, the
no-enhancement assumption is verified for values of $Q$ bigger than
or equal to a certain value $Q_{\text{min}}(\alpha,M,T)$. For
$\alpha\neq0$, the behavior of $Q_{\text{min}}$ is shown in
Fig.~\ref{TmaxQmin20}b for $M=20$ and for different transmissions.
We observe that it increases (step-wise, since only integer values
of the number of particles are permitted) as a function of the
decoherence rate for almost every transmission $T$, except for those
close to unity, for which it decreases. The behavior for small
values of $T$ can still be understood in terms of the average
current increase with $\alpha$. For $T=1$, being $Q_{\text{min}}=M$
at $\alpha=0$, decoherence can only cause a decrease.

\section{\label{sec:results}Results}

\begin{figure}
\includegraphics[width=8cm]{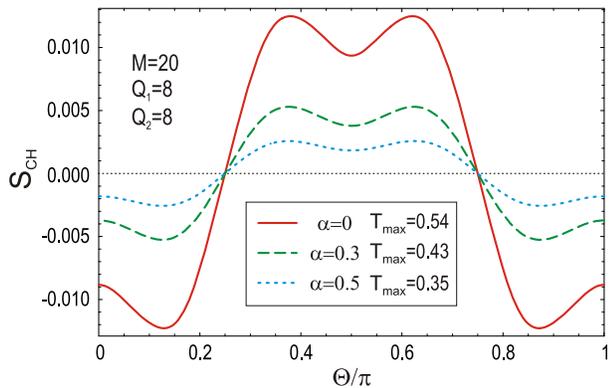}
\caption{(Color online) The quantity ${\cal S}_{\textrm{CH}}$ is
plotted as a function of $\Theta$ for $M=20$, $Q_1=Q_2=8$ and
different decoherence rates at the corresponding maximum allowed
transmissions. In particular, for $\alpha=0$ (solid line)
$T_{\text{max}}=0.54$, for $\alpha=0.3$ (dashed line)
$T_{\text{max}}=0.43$ and for  $\alpha=0.5$ (dotted line)
$T_{\text{max}}=0.35$. The amount of violation of the CH inequality
decreases with $\alpha$, whereas the range of angles for which
violation occurs does not change. We call $\Theta_{\text{best}}$ the
angle corresponding to the maximum violation.}\label{General20}
\end{figure}

In the present section we shall discuss how the CH inequality of
Eq.~(\ref{CHIreduced}) is affected by the presence of decoherence.
There are some general characteristics of the behavior of ${\cal
S}_{\textrm{CH}}$ that were already found in the absence of
decoherence \cite{faoro04} that hold also for finite $\alpha$
\cite{nota}. The most relevant are the following:
\begin{itemize}
\item ${\cal S}_{\textrm{CH}}$ is always symmetric as a function of $\Theta$ around
$\Theta=\pi/2$;
\item for given $M$, $Q_{1}$ and $Q_2$ the maximum violation always occurs for $T=T_{\text{max}}(\alpha,M,Q_1,Q_2)$.
\end{itemize}

In Fig.~\ref{General20}, ${\cal S}_{\textrm{CH}}$ is plotted as a
function of the angle $\Theta$ for $M=20$ and $Q_1=Q_2=8$. The three
curves refer, respectively, to $\alpha=0$ (solid line), $\alpha=0.3$
(dashed line) and  $\alpha=0.5$ (dotted line), each one calculated
for the corresponding $T=T_{\text{max}}$ reported in the label box.
The plot shows that the CH inequality is violated within a certain
window of values of $\Theta$. The violation is suppressed with
increasing decoherence rate, but occurs for the same range of
angles. This is due to the following properties of the joint
probabilities, which hold at $T=T_{\text{max}}$ for all values of
$\alpha$: i)
$P^{\pi/4}_{1,2}(Q_1,Q_2)=P^{3\pi/4}_{1,2}(Q_1,Q_2)=P_{1,-}(Q_1,Q_2)$,
as a consequence ${\cal S}_{\textrm{CH}}(\Theta=\pi/4,3\pi/4)=0$;
and ii) $P^{\Theta}_{1,2}(Q_1,Q_2)\ge P^{3\Theta}_{1,2}(Q_1,Q_2)$,
$P_{1,-}(Q_1,Q_2)$, $P_{-,2}(Q_1,Q_2)$ for $\pi/4\le \Theta\le
3\pi/4$. We checked that by reducing $T$ from $T_{\text{max}}$, but
keeping $\alpha$ constant, both the window of angles where violation
is present and its amount are decreased. Note that between
$\Theta=0$ and $\pi/2$, there is always an angle for which ${\cal
S}_{\textrm{CH}}$ is maximum, we shall denote it by
$\Theta_{\text{best}}(\alpha,M,Q_1,Q_2)$. For given $\alpha$, $M$,
$Q_1$ and $Q_2$, the maximum violation occurs at $T_{\text{max}}$
and $\Theta_{\text{best}}$.

\begin{figure}
\includegraphics[width=8.5cm]{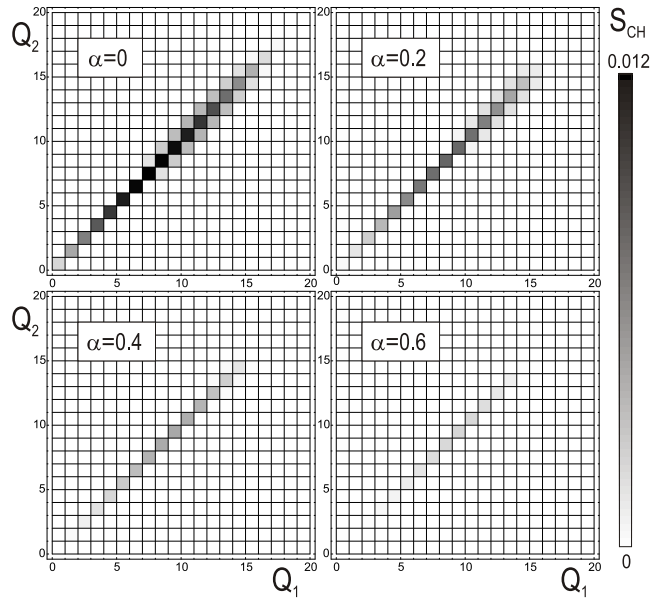}
\caption{Density plots of the maximum value of ${\cal
S}_{\textrm{CH}}$, evaluated at $T=T_{\text{max}}(\alpha,Q_1,Q_2)$
and $\Theta=\Theta_{\text{best}}(\alpha,Q_1,Q_2)$, in the
($Q_1,Q_2$) plane for $M=20$ relative to four different values of
decoherence ($\alpha=0$, $0.2$, $0.4$, $0.6$). ${\cal
S}_{\textrm{CH}}\simeq 0.012$ is the maximum violation for $M=20$
found in the absence of decoherence. }\label{DDP20}
\end{figure}
\begin{figure}
\includegraphics[width=8cm]{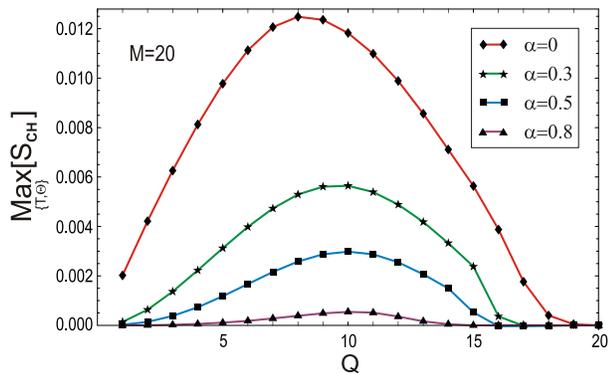}
\caption{(Color online) Maximum value of the quantity ${\cal
S}_{\textrm{CH}}$ ($T=T_{\text{max}}$ and
$\Theta=\Theta_{\text{best}}$) as a function of $Q$ for $M=20$ and
different values of decoherence: $\alpha=0$, $0.3$, $0.5$, $0.8$.
The  largest violation always occurs in the absence of decoherence
and, for a given $Q$, violation is reduced monotonically with
$\alpha$. The position where the maximum occurs, $Q_{\text{best}}$,
slightly increases with $\alpha$. At $Q=Q_{\text{best}}$,
$T_{\text{max}}=0.54$ for $\alpha=0$, $T_{\text{max}}=0.59$ for
$\alpha=0.3$, $T_{\text{max}}=0.52$ for $\alpha=0.5$ and
$T_{\text{max}}=0.39$ for $\alpha=0.8$. }\label{SmaxvsQ20}
\end{figure}

We now analyze the maximum violation of the CH inequality for a
given $M$ with $T=T_{\text{max}}$ and $\Theta=\Theta_{\text{best}}$
as a function of $Q_1, Q_2$ and $\alpha$. In Fig.~\ref{DDP20} we
show four density plots of ${\cal S}_{\textrm{CH}}$ in the
($Q_1,Q_2$) plane for different values of decoherence rate and
$M=20$. In the gray scale white corresponds to ${\cal
S}_{\textrm{CH}}=0$ and black to its maximum value taken in the
absence of decoherence. When $\alpha=0$, the CH inequality is
strongly violated for diagonal terms of the distribution (where
$Q_1=Q_2$). However, some weaker violations are also possible for
$Q_1\ne Q_2$, though they tend to disappear with increasing
$\alpha$.
By increasing the values of $\alpha$ the plots show that the maximum
violation of the CH inequality decreases rapidly: for $\alpha=0.6$
we get only $16\%$ of the largest value reached at $\alpha=0$. The
behavior of the CH inequality is symmetrical with respect to the
exchange of $Q_1$ with $Q_2$ for any rate of decoherence. In
Fig.~\ref{SmaxvsQ20} we report the section of the plots in
Fig.~\ref{DDP20} along the diagonal of the $(Q_1,Q_2)$-plane. The
four curves are relative to $\alpha=0$, 0.3, 0.5 and 0.8 an $M=20$.
Several observations are in order. If we denote with
$Q_{\text{best}}$ the position of the maximum of a curve, for all
values of decoherence rate $Q_{\text{best}}\sim M/2$, more
precisely, $Q_{\text{best}}=8$ for $\alpha=0$ and
$Q_{\text{best}}=10$ for all other curves. This slight increase of
$Q_{\text{best}}$ with $\alpha$ is due to the fact that an increase
in decoherence is accompanied by a slight enhancement of the average
current [Eq.~(\ref{i1})] flowing through the wires (as mentioned at
the end of Section \ref{Sec:system}). This is, however, a specific
feature of the model of decoherence we are considering. Note
furthermore that, as decoherence gets stronger, the range of values
of $Q$ for which violation takes place shrinks.

\begin{figure}
\includegraphics[width=8cm]{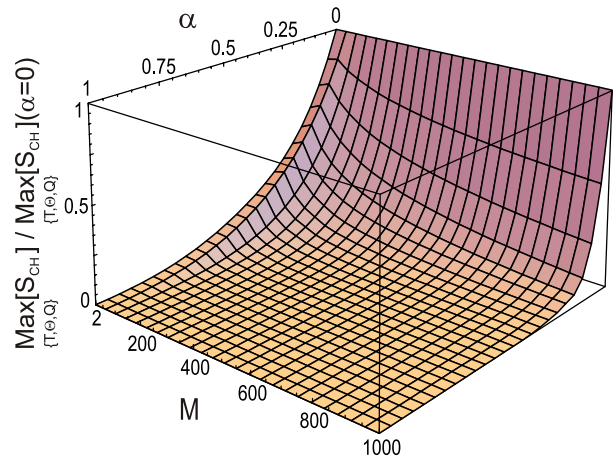}
\caption{(Color online) ${\cal S}_{\textrm{CH}}$, normalized to its
value in the absence of decoherence and calculated at
$T=T_{\text{max}}$, $\Theta=\Theta_{\text{best}}$ and
$Q=Q_{\text{best}}$, is plotted as a function of decoherence rate
$\alpha$ and number of injected entangled pairs $M$. Longer
measuring times (i.e. larger values of $M$) make decoherence more
effective, that is, make the detection of entanglement more
difficult. }\label{Smaxvsalpha}
\end{figure}

We now discuss the violation of the CH inequality as a function of
$\alpha$ and $M$ at $T=T_{\text{max}}$,
$\Theta=\Theta_{\text{best}}$ and $Q=Q_{\text{best}}$. In
Fig.~\ref{Smaxvsalpha} the ratio $s\equiv{\cal
S}_{\textrm{CH}}(\alpha,M,T_{\text{max}},\Theta_{\text{best}},Q_{\text{best}})/
{\cal
S}_{\textrm{CH}}(0,M,T_{\text{max}},\Theta_{\text{best}},Q_{\text{best}})$
(i. e. the quantity ${\cal S}_{\textrm{CH}}$ normalized to its value
in the absence of decoherence) is reported in a three-dimensional
plot as a function of $\alpha$ and the number of emitted pairs $M$.
The most interesting feature is that such a ratio decays more
rapidly with $\alpha$ as $M$ is increased. This means that
decoherence is more disruptive, as far as detection of entanglement
is concerned, for long measuring times (i.e. large $M$). As an
example, for $M=1000$ the extent of the violation is reduced by
$80\%$ at $\alpha=0.1$. More precisely, for values of $M$ larger
than $30$, we find that the normalized ${\cal S}_{\textrm{CH}}$
follows the law:
\begin{equation}
s\sim\frac{\sinh[K(1-\alpha)^{b \sqrt{M}}]}{\sinh(K)} ,
\end{equation}
with $K=7.26$ and $b=0.076$.

\begin{figure}
\includegraphics[width=8cm]{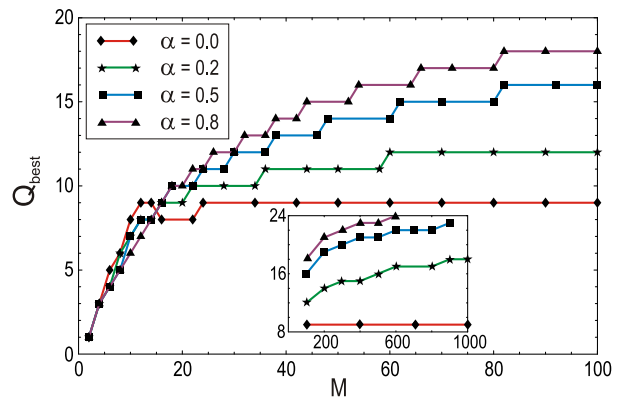}
\caption{(Color online) $Q_{\text{best}}$ is plotted as a function
of $M$ for different values of decoherence rate ($\alpha=0$, 0.2,
0.5 and 0.8). In the inset, curves are shown over an extended range,
up to $M=1000$. For a given $\alpha$, with increasing $M$ the value
of $Q_{\text{best}}$ increases very slowly remaining of the order of
10. }\label{QbestvsM}
\end{figure}

Another interesting aspect is related to $Q_{\text{best}}$ which, as
mentioned above, only slightly increases with $\alpha$ for all
values of $M$. As $M$ is increased, the value of $Q_{\text{best}}$,
for a given $\alpha$, does not increase proportionally to $M$, but
very much slowly and surprisingly remains of the order of 10 for
$M=1000$ (see Fig.~\ref{QbestvsM}). For $\alpha=0$ this can be
understood as follows. On the one hand, one expects
$Q_{\text{best}}$, corresponding to the largest ${\cal
S}_{\textrm{CH}}$, to be about the position of the maximum of joint
probability distributions, which can be assumed to be equal to the
product $MT$. On the other hand, $T_{\text{max}}$ is a decreasing
function of $M$, in fact it decays as $1/M$ \cite{faoro04}. The
product $MT_{\text{max}}$ is therefore expected to be a constant.
Indeed, it is possible to show, in the large $M$ expansion, that
$Q_{\text{best}}\sim MT_{\text{max}}$ for $\alpha=0$ and
$Q_{\text{best}}\sim M\sqrt{T_{\text{max}}}$ for $\alpha\neq 0$,
while $T_{\text{max}}\sim 1/M$ for $\alpha=0$ and
$T_{\text{max}}\sim 1/M^2$ for $\alpha\neq 0$. As a result,
$Q_{\text{best}}$ is roughly constant as a function of $M$ and
$\alpha$ \cite{nota0}.

\begin{figure}
\includegraphics[width=8cm]{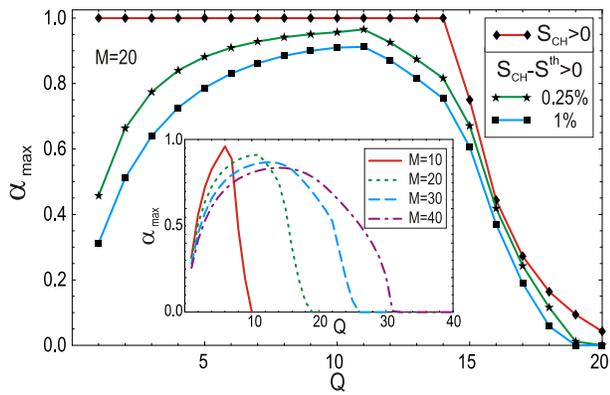}
\caption{(Color online) Maximum value ($\alpha_{\text{max}}$) of the
decoherence parameter for which there is still violation of the CH
inequality as a function of $Q$ for $M=20$. The line with
$\blacklozenge$ represents $\alpha_{\text{max}}(Q)$ with ${\cal
S}_{th}=0$: from $Q=1$ to $Q=14$ violation is found for any
decoherence rate. The line with $\bigstar$ is instead computed using
a threshold ${\cal S}_{th}$ which corresponds to $0.25\%$ of the
maximum violation value for $\alpha=0$, and the line with
$\blacksquare$ using ${\cal S}_{th}$ corresponding to $1\%$. The
latter threshold is used in the inset where $\alpha_{\text{max}}(Q)$
is plotted for $M=10$, $20$, $30$ and $40$. Interestingly we found
that there is a value $\bar{Q}$ that is more robust against
decoherence. In particular, $\bar{Q}=6$ for $M=10$, $\bar{Q}=11$ for
$M=20$, $\bar{Q}=13$ for $M=30$ and $\bar{Q}=14$ for $M=40$. With
increasing $M$, $\alpha_{\text{max}}$ diminishes slowly.
}\label{alphavsQ}
\end{figure}

The final point we address is the maximum decoherence rate,
that we denote by $\alpha_{\text{max}}$, for which there is still violation
of the CH inequality
as a function of $Q\equiv Q_1=Q_2$. In Fig.~\ref{alphavsQ} we plot
$\alpha_{\text{max}}$ as a function of $Q$ for $M=20$ at
$T=T_{\text{max}}$ and $\Theta=\Theta_{\text{best}}$. The line with
the symbol $\blacklozenge$ shows that violation of the CH inequality
is found for any rate of decoherence for $Q=1$ to $Q=14$ and
thereafter $\alpha_{\text{max}}$ decreases sharply. Nevertheless,
the extent of violation for $\alpha$ close to 1 is almost negligible
for most of $1\le Q\le 14$. One can therefore introduce a small
positive threshold ${\cal S}_{\text{th}}$ which defines the
violation as: ${\cal S}_{\textrm{CH}}<{\cal S}_{\text{th}}$. The
line with $\bigstar$ refers to a threshold of $0.25\%$ of the
maximum  value of ${\cal S}_{\textrm{CH}}$ at $\alpha=0$ (${\cal
S}_{\text{th}}=3\times 10^{-5}$ for $M=20$), and the line with
$\blacksquare$ to a $1\%$ (${\cal S}_{\text{th}}=1.2\times 10^{-4}$
for $M=20$). The latter percentage is used for the thresholds of the
plots in the inset of Fig.~\ref{alphavsQ}, where
$\alpha_{\text{max}}$ is plotted for $M=10$, $20$, $30$ and $40$. It
is shown that there are values of $Q$ that are more resistent to
decoherence, i. e. for which violation survives for larger
decoherence rates. In the caption of Fig.~\ref{alphavsQ}, the most
protected value against decoherence is denoted with $\bar{Q}\equiv
Q_{\text{best}}(\alpha_{\text{max}})$.

\section{\label{Sec:conclusions}Conclusions}

In this paper we have studied the effect of decoherence on the
violation of the CH inequality formulated in terms of the FCS
\cite{faoro04}. The system under investigation (Fig.~\ref{2VP})
consists of an idealized {\em entangler} connected, through a pair
of identical mesoscopic wires, to spin-selective counters. We have
assumed that decoherence, which occurs equally but independently in
the two conductors, is produced by the presence of additional
fictitious reservoirs according to the phenomenological model of
B\"uttiker \cite{buttiker86,buttiker88}. Decoherence is
parameterized by the rate $\alpha$.

As expected, decoherence gives rise to  suppression of the violation
of the CH inequality. The extent of such a suppression has been
analyzed as a function of the parameters which characterize the
system, namely the transmission $T$ of the wires, the angle between
analyzers $\Theta$, the number of injected entangled pairs $M$ and
the number of transmitted particles $Q_1$ and $Q_2$ in the counters.
First we have discussed the {\em no-enhancement assumption}, a
condition that needs to be satisfied in both leads 1 and 2 in order
for the CH inequality to hold. We have found that such condition, in
a given lead, is verified for all transmission $T$ up to some
maximum value $T_{\text{max}}$ which depends on $Q$, $M$ and, of course,
$\alpha$. In particular, $T_{\text{max}}$ decreases with the decoherence
rate up to some value of $Q$ and thereafter increases.
The main results can be summarized as follows:
\begin{itemize}
\item Maximal violation, even in the presence of
decoherence, occurs at the largest allowed transmission
$T=T_{\text{max}}$ and for $Q_1=Q_2$ (it disappears very rapidly
when $Q_1\neq Q_2$).
\item As long as $T=T_{\text{max}}$, the angle range of the analyzers for which violation takes place does not depend on
the decoherence rate, though the extent of violation decreases with
$\alpha$.
\item In the absence of decoherence, the maximum violation of the
CH inequality was proved to decay as $1/M$ \cite{faoro04}. Here we
have found that, for finite $\alpha$, the parameter ${\cal
S}_{\textrm{CH}}$ decreases exponentially with $\sqrt{M}$, more
precisely as $[f(\alpha)]^{\sqrt{M}}/M$, i.e. decays both with
increasing $M$ and $\alpha$.
\item The value of $Q$ for which maximum violation occurs is virtually independent of $M$, which means that the largest violations appear for relatively small numbers of transmitted particles, even at large observation times.
\item Interestingly, we have found that the largest
decoherence rate for which the CH inequality is violated (within a
given small tolerance) presents a maximum as a function
of $Q$. This means that there exist numbers of transmitted charges which are
more protected against decoherence, i.e. the influence of the
environment is less disruptive as far as the violation of CH
inequality is concerned.
\end{itemize}

Although, in this paper, dephasing is assumed to be produced by the
presence of additional reservoirs, other different sources of decoherence
are possible in mesoscopic systems. We believe that this model
captures the main effects of decoherence, as far as violations of
the CH inequality in a mesoscopic system is concerned, and that the
results found in this work may be useful to design the best
experimental conditions.

Since real systems cannot be perfectly shielded from the
environment, the issues analyzed in this work seem adequate not only
from a fundamental point of view, but also in what it might
contribute to the understanding of the properties of lossy quantum
channels. For the future it would be interesting to apply our method
to realistic systems, like normal or superconducting beam splitters.
Of interest would also be the combined effects of the presence of
spin-flipping processes and decoherence.

\begin{figure}
\includegraphics[width=8.5cm]{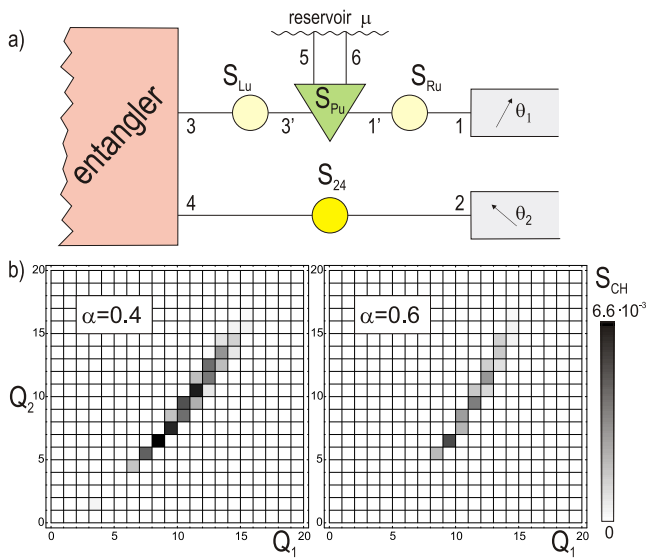}
\caption{(Color online) a) Idealized setup with a single additional
reservoir in the upper branch of the system. Scattering matrices are
chosen so that, in the absence of decoherence, the two conductors
have equal transmission. b) Density plot of the maximum value of
${\cal S}_{\textrm{CH}}$ in the ($Q_1,Q_2$)-plane for $M=20$ and for
$\alpha=0.4$ (left) and $\alpha=0.6$ (right). As decoherence
increases, the maximum violation is not achieved on the diagonal,
but is shifted towards the right-bottom part of the plane
($Q_1>Q_2$). This occurs because only the current flowing through
the conductor affected by decoherence is modified. Furthermore, the
suppression of the violation by $\alpha$ is less pronounced with
respect to the case with two additional reservoirs. For example, for
$\alpha=0$ we find $\textrm{Max}[{\cal S}_{\textrm{CH}}]=0.012$
achieved at $(Q_1=8,Q_2=8)$. For $\alpha=0.2$ and one additional
reservoir we have $\textrm{Max}[{\cal S}_{\textrm{CH}}]=0.0089$
reached at $(8,7)$, whereas for two additional reservoirs we get
$\textrm{Max}[{\cal S}_{\textrm{CH}}]=0.0074$ at $(9,9)$. For
$\alpha=0.4$ and one additional reservoir, $\textrm{Max}[{\cal
S}_{\textrm{CH}}]=0.0066$ at $(9,7)$, and with two additional
reservoirs, $\textrm{Max}[{\cal S}_{\textrm{CH}}]=0.0042$ at
$(10,10)$. Finally, for $\alpha=0.6$ and one additional reservoir,
$\textrm{Max}[{\cal S}_{\textrm{CH}}]=0.0046$ at $(10,7)$, and with
two additional reservoirs, $\textrm{Max}[{\cal
S}_{\textrm{CH}}]=0.0020$ at $(10,10)$. }\label{DP20setup}
\end{figure}

\begin{figure}
\includegraphics[width=8.5cm]{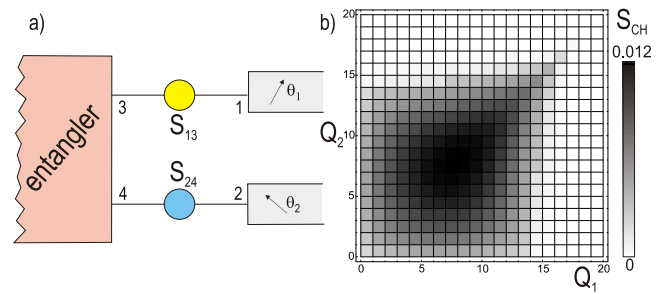}
\caption{(Color online) a) Idealized setup with no decoherence but
differently transmitting upper and lower conductors. b) The density
plot of the maximum value of the quantity ${\cal S}_{\textrm{CH}}$
is shown in the $(Q_1,Q_2)$-plane for $M=20$.}\label{DPTpTsetup}
\end{figure}

\begin{acknowledgments}
The authors would like to thank F. Sols for encouraging this work,
P. San-Jose for useful discussions and M. B\"{u}ttiker and C.W.J.
Beenakker for comments on the manuscript. This work was supported by
the FPI program of the Comunidad Aut\'onoma de Madrid, the EU under
the Marie Curie Research Training Network, EC-RTN Nano, EC-RTN
Spintronics and EC-IST-SQUIBIT2.
\end{acknowledgments}

\appendix

\section{Asymmetric setup: one additional reservoir}
\label{App:asym}

It is interesting to consider the case in which decoherence affects
the two wires differently. In this appendix we study the case when
decoherence affects only one of the two conductors, i.e. in the
presence of a single additional probe, for example, in the upper
branch, as depicted in Fig.~\ref{DP20setup}a. Being $T_0$ the
transmission of the elastic scatterers in the upper conductor, we
choose the transmission of the lower conductor to be equal to
$T=T_0^2/(2-T_0)^2$, in order for the two conductors to have the
same conductance in the absence of decoherence.
Fig.~\ref{DP20setup}b shows the density plot of the maximum value of
$S_{\textrm{CH}}$ (with $T=T_{\text{max}}$ and
$\Theta=\Theta_{\text{best}}$) as a function of $Q_1$ and $Q_2$, for
$M=20$, $\alpha=0.4$ (left) and $\alpha=0.6$ (right). In the
presence of decoherence the maximum violation is not achieved on the
diagonal ($Q_1=Q_2$), i.e. the behavior of $S_{\textrm{CH}}$ is not
symmetrical anymore with respect to $Q_1$ and $Q_2$. This is due to
the fact that, as we have seen above, the overall current increases
with $\alpha$ so that it is more likely to transmit a larger number
of particles in the conductor subjected to decoherence. Another
difference with respect to the case with two additional reservoirs
is that the suppression of the violation by $\alpha$ is less
pronounced.

The asymmetry found in the behavior of the density plots of
Fig.~\ref{DP20setup}b must not be confused with the asymmetry we
would obtain in a setup without decoherence but with conductors of
different resistance (system sketched in Fig.~\ref{DPTpTsetup}a). In
this case, by applying separately the no enhancement assumption to
the two conductors, large violations occur in a vast region of the
$(Q_1,Q_2)$-plane, as shown in Fig.~\ref{DPTpTsetup}b.
Interestingly, we note that one would get large violations of the CH
inequality for $Q_1\neq Q_2$. However, this asymmetry would not come
from the fact that entanglement is weakened by dephasing in only one
wire.

\begin{widetext}
\section{Expectation values}\label{Ap:expecvalues}

The most general expression for the characteristic function when
spin-$\sigma$ electrons are counted in lead 1 and spin-$\sigma'$
electrons are counted in lead 2 is
\begin{equation}\label{chiE2analyzers}
\chi_E (\lambda_{1\sigma},\lambda_{2\sigma'})=1+ \left(
e^{-i\lambda_{1\sigma}}-1 \right)\langle
\hat{N}_{\text{O}}^{1\sigma}\rangle + \left(
e^{-i\lambda_{2\sigma'}} -1 \right) \langle
\hat{N}_{\text{O}}^{2\sigma'} \rangle +\left(
e^{-i\lambda_{1\sigma}} -1 \right) \left( e^{-i\lambda_{2\sigma'}}
-1 \right) \langle \hat{N}_{\text{O}}^{1\sigma}
\hat{N}_{\text{O}}^{2\sigma'}\rangle
\end{equation}
for each relevant energy range: $0<E<\mu$ and $\mu<E<eV$. When both
spin species are counted in one of the terminals, the characteristic
function reads
\begin{eqnarray}\label{chiE1anlyzer}
\chi_E (\lambda_1,\lambda_{2\sigma'})&=&1+ \left( e^{-i\lambda_1} -1
\right) \langle ( \hat{N}_{\text{O}}^{1\uparrow} +
\hat{N}_{\text{O}}^{1\downarrow}) \rangle +\left(
e^{-i\lambda_{2\sigma'}} -1 \right) \langle
\hat{N}_{\text{O}}^{2\sigma'} \rangle + \left( e^{-i\lambda_1} -1
\right) \left( e^{-i\lambda_{2\sigma'}} -1 \right) \langle (
\hat{N}_{\text{O}}^{1\uparrow}+ \hat{N}_{\text{O}}^{1\downarrow})
\hat{N}_{\text{O}}^{2\sigma'} \rangle \nonumber\\&&+ \left(
e^{-i\lambda_1} -1 \right)^2 \langle \hat{N}_{\text{O}}^{1\uparrow}
\hat{N}_{\text{O}}^{1\downarrow} \rangle +\left( e^{-i\lambda_1} -1
\right)^2 \left( e^{-i\lambda_{2\sigma'}} -1 \right) \langle
\hat{N}_{\text{O}}^{1\uparrow} \hat{N}_{\text{O}}^{1\downarrow}
\hat{N}_{\text{O}}^{2\sigma'}\rangle
\end{eqnarray}
for counting both spins in terminal 1, where we have set $\lambda_{1\uparrow}=\lambda_{1\downarrow}\equiv \lambda_1$.

Using Eqs.~(\ref{counting}), (\ref{chiapproxSB}) and
(\ref{chiE2analyzers}), at zero temperature, one can calculate the
single terminal probability distribution:
\begin{eqnarray}\label{Ptheta1int}
P^{\theta_1}(Q_{1\sigma})=\frac{1}{2\pi}\int_{-\pi}^{\pi}d\lambda_{1\sigma}
\left[1+\left( e^{-i\lambda_{1\sigma}} -1 \right) \langle
\hat{N}_{\text{O}}^{1\sigma} \rangle
_S\right]^{M_{\mu}}\left[1+\left( e^{-i\lambda_{1\sigma}} -1 \right)
\langle \hat{N}_{\text{O}}^{1\sigma} \rangle
_B\right]^{M-M_{\mu}}e^{i\lambda_{1\sigma}Q_{1\sigma}},
\end{eqnarray}
where $\langle \hat{N}_{\text{O}}^{1\sigma} \rangle
_{S,B}\equiv\bra{\psi_{S,B}}
\hat{N}_{\text{O}}^{1\sigma}\ket{\psi_{S,B}}$.
After integration over $\lambda_{1\sigma}$ we get
\begin{eqnarray}\label{Ptheta1sum}
P^{\theta_1}(Q_{1\sigma})&=&\left(1-\langle\hat{N}_{\text{O}}^{1\sigma}
\rangle
_S\right)^{M_{\mu}}\left(1-\langle\hat{N}_{\text{O}}^{1\sigma}\rangle
_B\right)^{M-M_{\mu}-Q_{1\sigma}}\left(\langle\hat{N}_{\text{O}}^{1\sigma}\rangle
_B\right)^{Q_{1\sigma}}\nonumber\\&\times&
\sum_{n=\text{Max}[0,Q_{1\sigma}-M+M_{\mu}]}^{\text{Min}[M_{\mu},Q_{1\sigma}]}
\left(\begin{array}{c}M_{\mu}\\n
\end{array}\right)\left(\begin{array}{c}M-M_{\mu}\\Q_{1\sigma}-n
\end{array}\right)\left[\frac{\langle\hat{N}_{\text{O}}^{1\sigma}
\rangle _S(1-\langle\hat{N}_{\text{O}}^{1\sigma}\rangle
_B)}{\langle\hat{N}_{\text{O}}^{1\sigma} \rangle
_B(1-\langle\hat{N}_{\text{O}}^{1\sigma}\rangle _S)}\right]^n.
\end{eqnarray}
If one chooses $\mu_{\text{L}}=eV$ and $\mu_{\text{R}}=0$, one
obtains $\mu=\tfrac{eV}{2}$ and $M_{\mu}=\tfrac{M}{2}$. Therefore,
in order for $M_{\mu}$ to be integer, $M$ must be an even number.

Using Eqs.~(\ref{counting}), (\ref{chiapproxSB}) and
(\ref{chiE1anlyzer}), the single terminal probability distribution
when both spin species are counted in the terminal is
\begin{eqnarray}
P(Q_{1})=\frac{1}{2\pi}\int_{-\pi}^{\pi}\left[1+\left(
e^{-i\lambda_{1}} -1 \right)\langle\hat{N}_{\text{O}}^{1} \rangle
_S+\left( e^{-i\lambda_{1}} -1
\right)^2\langle\hat{N}_{\text{O}}^{1\uparrow}\hat{N}_{\text{O}}^{1\downarrow}\rangle
_S\right]^{M_{\mu}}\nonumber \\\times\left[1+\left(
e^{-i\lambda_{1}} -1 \right)\langle\hat{N}_{\text{O}}^{1} \rangle
_B+\left( e^{-i\lambda_{1}} -1
\right)^2\langle\hat{N}_{\text{O}}^{1\uparrow}\hat{N}_{\text{O}}^{1\downarrow}\rangle
_B\right]^{M-M_{\mu}}e^{i\lambda_{1}Q_{1}}d\lambda_{1},
\end{eqnarray}
where $\langle\hat{N}_{\text{O}}^{1} \rangle_{S,B}\equiv\langle\left(\hat{N}_{\text{O}}^{1\uparrow}+\hat{N}_{\text{O}}^{1\downarrow}\right)\rangle_{S,B}$.

In the following subsections we collect all the expectation values needed to work
out the expressions for the probability distributions of Eq.~(\ref{CHIreduced}).

\subsection{Setup with two additional reservoirs}

Let us consider the setup depicted in Fig.~\ref{2VP}, where two
additional reservoirs are present. The expectation values for energies
$0<E<\mu$ as a function of transmission $T$, decoherence parameter
$\alpha$ and analyzers' angle $\Theta$ are
\begin{eqnarray}
\langle \hat{N}_{\text{O}}^{1\uparrow} \rangle _{S}&=&\langle
\hat{N}_{\text{O}}^{2\uparrow} \rangle
_{S}=\frac{2\sqrt{T}[\sqrt{T}+\alpha(2-\sqrt{T})-\alpha^2(1-\sqrt{T})]}{[2-\alpha(1-\sqrt{T})]^2},\\
\langle \hat{N}_{\text{O}}^{1} \rangle _{S}&=&\langle
\hat{N}_{\text{O}}^{2} \rangle
_{S}=\frac{4\sqrt{T}[\sqrt{T}+\alpha(2-\sqrt{T})-\alpha^2(1-\sqrt{T})]}{[2-\alpha(1-\sqrt{T})]^2},\\
\langle
\hat{N}_{\text{O}}^{1\uparrow}\hat{N}_{\text{O}}^{1\downarrow}
\rangle _{S}=\langle
\hat{N}_{\text{O}}^{2\uparrow}\hat{N}_{\text{O}}^{2\downarrow}
\rangle _{S}&=&\frac{4T\alpha[2\sqrt{T}+2\alpha(1-\sqrt{T})-\alpha^2(1-\sqrt{T})]}{[2-\alpha(1-\sqrt{T})]^3},\\
\langle \hat{N}_{\text{O}}^{1\uparrow}\hat{N}_{\text{O}}^{2\uparrow}
\rangle _{S}&=&\frac{4T\{[\alpha(2-\alpha)+\sqrt{T}(1-\alpha+\alpha^2)]^2-T(1-\alpha)^2\cos\Theta\}}{[2-\alpha(1-\sqrt{T})]^4},\\
\langle \hat{N}_{\text{O}}^{1}\hat{N}_{\text{O}}^{2\uparrow} \rangle
_{S}=\langle \hat{N}_{\text{O}}^{1\uparrow}\hat{N}_{\text{O}}^{2}
\rangle _{S}&=&\frac{8T[\sqrt{T}+\alpha(2-\sqrt{T})-\alpha^2(1-\sqrt{T})]^2}{[2-\alpha(1-\sqrt{T})]^4},\\
\langle
\hat{N}_{\text{O}}^{1\uparrow}\hat{N}_{\text{O}}^{1\downarrow}\hat{N}_{\text{O}}^{2\uparrow}
\rangle _{S}=\langle
\hat{N}_{\text{O}}^{1\uparrow}\hat{N}_{\text{O}}^{2\uparrow}\hat{N}_{\text{O}}^{2\downarrow}
\rangle
_{S}&=&8\sqrt{T^3}\alpha[2\sqrt{T}+2\alpha(1-\sqrt{T})-\alpha^2(1-\sqrt{T})]\nonumber\\&&
\times\frac{[\sqrt{T}+\alpha(2-\sqrt{T})-\alpha^2(1-\sqrt{T})]}{[2-\alpha(1-\sqrt{T})]^5}.
\end{eqnarray}
The expectation values for energies $\mu<E<eV$ are
\begin{eqnarray}
\langle \hat{N}_{\text{O}}^{1\uparrow} \rangle _{B}&=&\langle
\hat{N}_{\text{O}}^{2\uparrow} \rangle
_{B}=\frac{2T(1-\alpha)}{[2-\alpha(1-\sqrt{T})]^2},\\
\langle \hat{N}_{\text{O}}^{1} \rangle _{B}&=&\langle
\hat{N}_{\text{O}}^{2} \rangle
_{B}=\frac{4T(1-\alpha)}{[2-\alpha(1-\sqrt{T})]^2},\\
\langle
\hat{N}_{\text{O}}^{1\uparrow}\hat{N}_{\text{O}}^{1\downarrow}
\rangle _{B}&=&\langle
\hat{N}_{\text{O}}^{2\uparrow}\hat{N}_{\text{O}}^{2\downarrow}
\rangle _{B}=0,\\
\langle \hat{N}_{\text{O}}^{1\uparrow}\hat{N}_{\text{O}}^{2\uparrow}
\rangle _{B}&=&\frac{8T^2(1-\alpha)^2\sin^2(\Theta/2)}{[2-\alpha(1-\sqrt{T})]^4},\\
\langle \hat{N}_{\text{O}}^{1}\hat{N}_{\text{O}}^{2\uparrow} \rangle
_{B}&=&\langle \hat{N}_{\text{O}}^{1\uparrow}\hat{N}_{\text{O}}^{2}
\rangle _{B}=\frac{8T^2(1-\alpha)^2}{[2-\alpha(1-\sqrt{T})]^4},\\
\langle
\hat{N}_{\text{O}}^{1\uparrow}\hat{N}_{\text{O}}^{1\downarrow}\hat{N}_{\text{O}}^{2\uparrow}
\rangle _{B}&=&\langle
\hat{N}_{\text{O}}^{1\uparrow}\hat{N}_{\text{O}}^{2\uparrow}\hat{N}_{\text{O}}^{2\downarrow}
\rangle _{B}=0.
\end{eqnarray}

\subsection{Setup with one additional reservoir}

Let us now consider the asymmetrical setup of Fig.~\ref{DP20setup},
where there is only one additional reservoir. For energies $0<E<\mu$ we
have that $\langle \hat{N}_{\text{O}}^{1\uparrow} \rangle _{S}$,
$\langle \hat{N}_{\text{O}}^{1} \rangle _{S}$ and
$\langle\hat{N}_{\text{O}}^{1\uparrow}\hat{N}_{\text{O}}^{1\downarrow}\rangle
_{S}$ are equal to the case with two fictitious reservoirs. The other
expectation values are
\begin{eqnarray}
\langle \hat{N}_{\text{O}}^{2\uparrow} \rangle
_{S}&=&\frac{T}{2},\\
\langle \hat{N}_{\text{O}}^{2} \rangle
_{S}&=&T,\\
\langle
\hat{N}_{\text{O}}^{2\uparrow}\hat{N}_{\text{O}}^{2\downarrow}
\rangle _{S}&=&0,\\
\langle \hat{N}_{\text{O}}^{1\uparrow}\hat{N}_{\text{O}}^{2\uparrow}
\rangle _{S}&=&\frac{\sqrt{T^3}\alpha(2-\alpha)+T^2[1-\alpha+\alpha^2-(1-\alpha)\cos\Theta]}{[2-\alpha(1-\sqrt{T})]^2},\\
\langle \hat{N}_{\text{O}}^{1}\hat{N}_{\text{O}}^{2\uparrow} \rangle
_{S}&=&\langle \hat{N}_{\text{O}}^{1\uparrow}\hat{N}_{\text{O}}^{2}
\rangle _{S}=\frac{2\sqrt{T^3}[\sqrt{T}+\alpha(2-\sqrt{T})-\alpha^2(1-\sqrt{T})]}{[2-\alpha(1-\sqrt{T})]^4},\\
\langle
\hat{N}_{\text{O}}^{1\uparrow}\hat{N}_{\text{O}}^{1\downarrow}\hat{N}_{\text{O}}^{2\uparrow}
\rangle _{S}&=&\frac{2T^2\alpha[2\sqrt{T}+2\alpha(1-\sqrt{T})-\alpha^2(1-\sqrt{T})]}{[2-\alpha(1-\sqrt{T})]^3},\\
\langle
\hat{N}_{\text{O}}^{1\uparrow}\hat{N}_{\text{O}}^{2\uparrow}\hat{N}_{\text{O}}^{2\downarrow}
\rangle _{S}&=&0.
\end{eqnarray}
For energies $\mu<E<eV$ we have that $\langle
\hat{N}_{\text{O}}^{1\uparrow} \rangle _{B}$, $\langle
\hat{N}_{\text{O}}^{1} \rangle _{B}$, $\langle
\hat{N}_{\text{O}}^{1\uparrow}\hat{N}_{\text{O}}^{1\downarrow}
\rangle _{B}$, $\langle
\hat{N}_{\text{O}}^{2\uparrow}\hat{N}_{\text{O}}^{2\downarrow}
\rangle _{B}$, $\langle
\hat{N}_{\text{O}}^{1\uparrow}\hat{N}_{\text{O}}^{1\downarrow}\hat{N}_{\text{O}}^{2\uparrow}
\rangle _{B}$ and $\langle
\hat{N}_{\text{O}}^{1\uparrow}\hat{N}_{\text{O}}^{2\uparrow}\hat{N}_{\text{O}}^{2\downarrow}
\rangle _{B}$ are equal to the case with two additional reservoirs.
The other expectation values are
\begin{eqnarray}
\langle \hat{N}_{\text{O}}^{2\uparrow} \rangle
_{B}&=&\frac{T}{2},\\
\langle \hat{N}_{\text{O}}^{2} \rangle
_{B}&=&T,\\
\langle \hat{N}_{\text{O}}^{1\uparrow}\hat{N}_{\text{O}}^{2\uparrow}
\rangle _{B}&=&\frac{2T^2(1-\alpha)\sin^2(\Theta/2)}{[2-\alpha(1-\sqrt{T})]^2},\\
\langle \hat{N}_{\text{O}}^{1}\hat{N}_{\text{O}}^{2\uparrow} \rangle
_{B}&=&\langle \hat{N}_{\text{O}}^{1\uparrow}\hat{N}_{\text{O}}^{2}
\rangle _{B}=\frac{2T^2(1-\alpha)}{[2-\alpha(1-\sqrt{T})]^2}.
\end{eqnarray}

\section{Probability distributions}
\label{Ap:probdistributions}

In order to calculate the various probabilities needed to evaluate
the CH inequality, Eq.~(\ref{CHI}) and (\ref{CHIreduced}), it is
necessary to solve the integrals of Eq.~(\ref{counting}), where the
different characteristic functions are given in
Eq.~(\ref{chiE2analyzers}) and Eq.~(\ref{chiE1anlyzer}). As we
mentioned in Section \ref{Sec:CHI}, explicit expressions of the
probability distributions in terms of sums are lengthy and
complicated for practical calculations. It is possible,
nevertheless,  to express the result for the various probabilities
in a quite simple fashion, which makes them manageable for
computational analysis. The point is to realize that the
characteristic functions are nothing but polynomial functions on the
variables $e^{i\lambda_i}$ of different degrees. The effect of each
integral of Eq.~(\ref{counting}), together with its accompanying
complex exponential $e^{i\lambda_iQ_i}/(2\pi)$, is simply to select
the coefficient of the characteristic function polynomial which
corresponds to the power equal to $Q_i$. Consequently, the result of
the integrals for the various probabilities can be expressed in
terms of $Q_i$ order derivatives of the characteristic function, as
we show below. Note that, whenever both analyzers are present, we
will choose without loss of generality: $\sigma=\uparrow$ and
$\sigma'=\uparrow$. For single terminal probability distributions we
will also set $\sigma=\uparrow$ for lead 1 and $\sigma'=\uparrow$
for lead 2, although resulting expressions will not depend either on
the direction of the spin or on the angle of the analyzer.

For the single terminal probability distribution with analyzer we
have
\begin{equation}
P^{\theta_1}(Q_{1\uparrow})=\frac{1}{2\pi}\int_{-\pi}^{\pi}d\lambda_{1\uparrow}e^{i\lambda_{1\uparrow}Q_{1\uparrow}}\chi(\lambda_{1\uparrow})=
\frac{1}{Q_{1\uparrow}!}\left.\frac{d^{Q_{1\uparrow}}\chi(\lambda_{1\uparrow})}{d(e^{i\lambda_{1\uparrow}})^{Q_{1\uparrow}}}
\right|_{e^{i\lambda_{1\uparrow}}\rightarrow 0},
\end{equation}
where $\chi(\lambda_{1\uparrow})$ can be extracted from Eq.
(\ref{chiE2analyzers}) making $\lambda_{2\uparrow}=0$ and using Eq.
(\ref{chiapproxSB}).
Since $\mu=\tfrac{eV}{2}$, we have that $M_{\mu}=\tfrac{M}{2}$,
being $M$ the total number of emitted particles per lead or per
spin. The expectation values needed in Eq.~(\ref{chiE2analyzers})
above and below energy $\mu$ are given in Appendix
\ref{Ap:expecvalues}. The single terminal probability distribution
in the absence of analyzer is
\begin{equation}
P(Q_{1})=\frac{1}{2\pi}\int_{-\pi}^{\pi}d\lambda_{1}e^{i\lambda_{1}Q_{1}}\chi(\lambda_{1})=
\frac{1}{Q_{1}!}\left.\frac{d^{Q_{1}}\chi(\lambda_{1})}{d(e^{i\lambda_{1}})^{Q_{1}}}\right|_{e^{i\lambda_{1}}\rightarrow
0},
\end{equation}
where the characteristic function can be extracted now from Eq.~(\ref{chiE1anlyzer}), setting again $\lambda_{2\uparrow}=0$. We can
get similarly the expressions for $P^{\theta_2}(Q_{2\uparrow})$ and
$P(Q_{2})$.

The joint probability distribution when both analyzers are present
gives
\begin{eqnarray}
P^{\theta_1,\theta_2}(Q_{1\uparrow},Q_{2\uparrow})&=&\frac{1}{(2\pi)^2}\int_{-\pi}^{\pi}d\lambda_{1\uparrow}e^{i\lambda_{1\uparrow}Q_{1\uparrow}}
\int_{-\pi}^{\pi}d\lambda_{2\uparrow}e^{i\lambda_{2\uparrow}Q_{2\uparrow}}\chi(\lambda_{1\uparrow},\lambda_{2\uparrow})
\nonumber\\&=&\frac{1}{Q_{1\uparrow}!Q_{2\uparrow}!}\left.\frac{d^{Q_{1\uparrow}}d^{Q_{2\uparrow}}\chi(\lambda_{1\uparrow},\lambda_{2\uparrow})}
{d(e^{i\lambda_{1\uparrow}})^{Q_{1\uparrow}}d(e^{i\lambda_{2\uparrow}})^{Q_{2\uparrow}}}
\right|_{e^{i\lambda_{1\uparrow}},e^{i\lambda_{2\uparrow}}\rightarrow
0},
\end{eqnarray}
which only depends on the angle
$\Theta\equiv(\theta_1\pm\theta_2)/2$, as we showed in Section
\ref{Sec:CHI}. When there is only one analyzer we have
\begin{eqnarray}
P^{\theta_1,-}(Q_{1\uparrow},Q_{2})=\frac{1}{(2\pi)^2}\int_{-\pi}^{\pi}d\lambda_{1\uparrow}e^{i\lambda_{1\uparrow}Q_{1\uparrow}}
\int_{-\pi}^{\pi}d\lambda_{2}e^{i\lambda_{2}Q_{2}}\chi(\lambda_{1\uparrow},\lambda_{2})
=\frac{1}{Q_{1\uparrow}!Q_{2}!}\left.\frac{d^{Q_{1\uparrow}}d^{Q_{2}}\chi(\lambda_{1\uparrow},\lambda_{2})}
{d(e^{i\lambda_{1\uparrow}})^{Q_{1\uparrow}}d(e^{i\lambda_{2}})^{Q_{2}}}
\right|_{e^{i\lambda_{1\uparrow}},e^{i\lambda_{2}}\rightarrow 0}
\end{eqnarray}
and
\begin{eqnarray}
P^{-,\theta_2}(Q_{1},Q_{2\uparrow})=\frac{1}{(2\pi)^2}\int_{-\pi}^{\pi}d\lambda_{1}e^{i\lambda_{1}Q_{1}}
\int_{-\pi}^{\pi}d\lambda_{2\uparrow}e^{i\lambda_{2\uparrow}Q_{2\uparrow}}\chi(\lambda_{1},\lambda_{2\uparrow})
=\frac{1}{Q_{1}!Q_{2\uparrow}!}\left.\frac{d^{Q_{1}}d^{Q_{2\uparrow}}\chi(\lambda_{1},\lambda_{2\uparrow})}
{d(e^{i\lambda_{1}})^{Q_{1}}d(e^{i\lambda_{2\uparrow}})^{Q_{2\uparrow}}}
\right|_{e^{i\lambda_{1}},e^{i\lambda_{2\uparrow}}\rightarrow 0}.
\end{eqnarray}
However, this two last expressions are not strictly needed since one can use the relations in Eq.~(\ref{factorized}).
Again, all the expectation values which are needed for evaluating
these probabilities are given in Appendix \ref{Ap:expecvalues}.
\end{widetext}


\begin{thebibliography}{99}
\bibitem{newnota1} Note that we make use of the CH inequality
as a tool to detect the persistence of entanglement in the presence
of decoherence processes, given an injected entangled state (whose
existence is assumed).
\bibitem{nota1} Very recently an alternative to B\"uttiker's dephasing model was put forward by Beenakker \cite{beenakker05}: the fictitious lead is replaced by a fictitious stub. It turns out that voltage fluctuations in the stub suppress quantum interference effects without affecting the noise properties.
\bibitem{nota} In particular, if the entangler is substituted with a source that emits factorized states, the CH inequality is never violated for any decoherence rate. This is simply due to the fact that two terminal joint probability distributions are given by the product of single terminal probability distributions, thus making the CH inequality always equal to zero.
\bibitem{nota0} More precisely, for large $M$, $Q_{\text{best}}$ is implicitly given by the equation: $Q_{\text{best}}=\tilde{Q}[\alpha,T_{\text{max}}(\alpha,M,Q_{\text{best}})]$,
where $\tilde{Q}[\alpha,T]=M[\sqrt{T}\alpha(2-\alpha)+T(2-2\alpha+\alpha^2)]/[2+(\sqrt{T}-1)\alpha]^2$ is the position of the maximum of the distribution in the large $M$ approximation.
\end{thebibliography}

\end{document}